\documentclass[sigconf]{acmart}

\AtBeginDocument{%
  \providecommand\BibTeX{{%
    \normalfont B\kern-0.5em{\scshape i\kern-0.25em b}\kern-0.8em\TeX}}}

\usepackage{multirow}

\usepackage{graphicx}  %
\usepackage{algorithm}%
\usepackage{algpseudocode}%
\usepackage{subcaption}
\usepackage{amssymb}
\usepackage{enumitem}
\usepackage[stable]{footmisc}

\copyrightyear{2020} 
\acmYear{2020} 
\setcopyright{acmlicensed}\acmConference[KDD '20]{Proceedings of the 26th ACM SIGKDD Conference on Knowledge Discovery and Data Mining}{August 23--27, 2020}{Virtual Event, CA, USA}
\acmBooktitle{Proceedings of the 26th ACM SIGKDD Conference on Knowledge Discovery and Data Mining (KDD '20), August 23--27, 2020, Virtual Event, CA, USA}
\acmPrice{15.00}
\acmDOI{10.1145/3394486.3403249}
\acmISBN{978-1-4503-7998-4/20/08}

\begin{document}
\fancyhead{}

\title{DeepSinger: Singing Voice Synthesis with Data Mined \\From the Web}

\author{Yi Ren$^{1*}$, Xu Tan$^{2*}$, Tao Qin$^{2}$, Jian Luan$^{3}$, Zhou Zhao$^{1\dagger}$, Tie-Yan Liu$^{2}$}
\affiliation{$^{1}$Zhejiang University, $^{2}$Microsoft Research Asia, $^{3}$Microsoft STC Asia}
\email{rayeren@zju.edu.cn,{xuta,taoqin,jianluan}@microsoft.com,zhaozhou@zju.edu.cn,tyliu@microsoft.com}

\renewcommand{\shortauthors}{Ren and Tan, et al.}

\begin{abstract}
In this paper\footnote{$^*$ Equal contribution. $\dagger$ Corresponding author.}, we develop DeepSinger, a multi-lingual multi-singer singing voice synthesis (SVS) system, which is built from scratch using singing training data mined from music websites. The pipeline of DeepSinger consists of several steps, including data crawling, singing and accompaniment separation, lyrics-to-singing alignment, data filtration, and singing modeling. Specifically, we design a lyrics-to-singing alignment model to automatically extract the duration of each phoneme in lyrics starting from coarse-grained sentence level to fine-grained phoneme level, and further design a multi-lingual multi-singer singing model based on a feed-forward Transformer to directly generate linear-spectrograms from lyrics, and synthesize voices using Griffin-Lim. DeepSinger has several advantages over previous SVS systems: 1) to the best of our knowledge, it is the first SVS system that directly mines training data from music websites, 2) the lyrics-to-singing alignment model further avoids any human efforts for alignment labeling and greatly reduces labeling cost, 3) the singing model based on a feed-forward Transformer is simple and efficient, by removing the complicated acoustic feature modeling in parametric synthesis and leveraging a reference encoder to capture the timbre of a singer from noisy singing data, and 4) it can synthesize singing voices in multiple languages and multiple singers. We evaluate DeepSinger on our mined singing dataset that consists of about 92 hours data from 89 singers on three languages (Chinese, Cantonese and English). The results demonstrate that with the singing data purely mined from the Web, DeepSinger can synthesize high-quality singing voices in terms of both pitch accuracy and voice naturalness\footnote{Our audio samples are shown in \url{https://speechresearch.github.io/deepsinger/}.}. 
\end{abstract}



\maketitle

\section{Introduction}
Singing voice synthesis (SVS)~\cite{nishimura2016singing,blaauw2017neural,lee2019adversarially,lu2020xiaoicesing}, which generates singing voices from lyrics, has attracted a lot of attention in both research and industrial community in recent years. Similar to text to speech (TTS)~\cite{shen2018natural,ren2019fastspeech,ren2020fastspeech} that enables machines to speak, SVS enables machines to sing, both of which have been greatly improved with the rapid development of deep neural networks. Singing voices have more complicated prosody than normal speaking voices~\cite{umbert2015expression}, and therefore SVS needs additional information to control the duration and the pitch of singing voices, which makes SVS more challenging than TTS~\cite{nguyen2018study}.

Previous works on SVS include lyrics-to-singing alignment~\cite{fujihara2011lyricsynchronizer,chien2016alignment,gupta2018semi}, parametric synthesis~\cite{kim2018korean,blaauw2017neural}, acoustic modeling~\cite{nishimura2016singing,nakamura2019singing,lu2020xiaoicesing}, and adversarial synthesis~\cite{chandna2019wgansing,lee2019adversarially,hono2019singing}. Although they achieve reasonably good performance, these systems typically require 1) a large amount of high-quality singing recordings as training data, and 2) strict data alignments between lyrics and singing audio for accurate singing modeling, both of which incur considerable data labeling cost. Previous works collect these two kinds of data as follows:
\begin{itemize}[leftmargin=*]
\item For the first kind of data, previous works usually ask human to sing songs in a professional recording studio to record high-quality singing voices for at least a few hours, which needs the involvement of human experts and is costly. 
\item For the second kind of data, previous works usually first manually split a whole song into aligned lyrics and audio in sentence level~\cite{lee2019adversarially}, and then extract the duration of each phoneme either by manual alignment or a phonetic timing model~\cite{nishimura2016singing,blaauw2017neural}, which incurs data labeling cost. 
\end{itemize}

As can be seen, the training data for SVS mostly rely on human recording and annotations. What is more, there are few publicly available singing datasets, which increases the entry cost for researchers to work on SVS and slows down the research and product application in this area. Considering a lot of tasks such as language modeling and generation, search and ads ranking, and image classification heavily rely on data collected from the Web, a natural question is that: Can we build an SVS system with data collected from the Web? While there are plenty of songs in music websites and mining training data from the Web seems promising, we face several technical challenges:
\begin{itemize}[leftmargin=*]
\item The songs in music websites usually mix singing with accompaniment, which are very noisy to train SVS systems. In order to leverage such data, singing and accompaniment separation is needed to obtain clean singing voices.  
\item The singing and the corresponding lyrics in crawled songs are not always well matched due to possible errors in music websites. How to filter the mismatched songs is important to ensure the quality of the mined dataset.  
\item Although some songs have time alignments between the singing and lyrics in sentence level, most alignments are not accurate. For accurate singing modeling, we need build our own alignment model for both sentence-level alignment and phoneme-level alignment.  
\item There are still noises in the singing audio after separation. How to design a singing model to learn from noisy data is challenging.
\end{itemize}

In this paper, we develop DeepSinger, a singing voice synthesis system that is built from scratch by using singing training data mined from music websites. To address the above challenges, we design a pipeline in DeepSinger that consists of several data mining and modeling steps, including:
\begin{itemize}[leftmargin=*]
    \item Data crawling. We crawl popular songs of top singers in multiple languages from a music website.
    \item Singing and accompaniment separation. We use a popular music separation tool Spleeter~\cite{hennequin2019spleeter} to separate singing voices from song accompaniments.
    \item Lyrics-to-singing alignment. We build an alignment model to segment the audio into sentences and extract the singing duration of each phoneme in lyrics. 
    \item Data filtration. We filter the aligned lyrics and singing voices according to their confidence scores in alignment.
    \item Singing modeling. We build a FastSpeech~\citep{ren2019fastspeech,ren2020fastspeech} based singing model and leverage a reference encoder to handle noisy data.
\end{itemize}
Specifically, the detailed designs of the lyrics-to-singing alignment model and singing model are as follows:
\begin{itemize}[leftmargin=*]
    \item We build the lyrics-to-singing alignment model based on automatic speech recognition to extract the duration of each phoneme in lyrics starting from coarse-grained sentence level to fine-grained phoneme level. 
    \item We design a multi-lingual multi-singer singing model based on FastSpeech~\cite{ren2019fastspeech,ren2020fastspeech} to directly generate linear-spectrogram from lyrics, instead of the traditional acoustic features in parametric synthesis. Additionally, we design a reference encoder in the singing model to capture the timbre of a singer from noisy singing data, instead of using singer ID.
\end{itemize}

We conduct experiments on our mined singing dataset (92 hours data with 89 singers and three languages) to evaluate the effectiveness of DeepSinger. Experiment results show that with the singing data purely mined from the Web, DeepSinger can synthesize high-quality singing voices in terms of both pitch accuracy and voice naturalness. 

The contributions of this paper are summarized as follows:
\begin{itemize}[leftmargin=*]
    \item To the best of our knowledge, DeepSinger is the first SVS system built from data directly mined from the Web, without any high-quality singing data recorded by human.
    \item The lyrics-to-singing alignment model avoids any human efforts for alignment labeling and greatly reduces labeling cost.
    \item The FastSpeech based singing model is simple and efficient, by removing the complicated acoustic feature modeling in parametric synthesis and leveraging a reference encoder to capture the timbre of a singer from noisy singing data.
    \item DeepSinger can synthesize high-quality singing voices in multiple languages and multiple singers. 
\end{itemize}

\section{Background}

\begin{figure*}[!htb] 
	\centering
	\includegraphics[width=\textwidth,trim={0 2.0cm 0 0},clip]{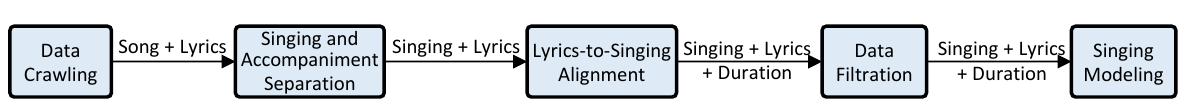}
	\caption{Overview of the DeepSinger pipeline.}
	\label{arch_pipeline}
	\vspace{-0.4cm}
\end{figure*}

In this section, we introduce the background of DeepSinger, including text to speech (TTS), singing voice synthesis (SVS), text-to-audio alignment that is the key component to SVS system, as well as some other works that leverage training data mined from the Web.

\paragraph{Text to Speech}
Text to Speech (TTS)~\cite{shen2018natural,ren2019fastspeech,ren2020fastspeech} aims to synthesize natural and intelligible speech given text as input, which has witnessed great progress in recent years. TTS systems have changed from concatenative synthesis~\cite{hunt1996unit}, to statistical parametric synthesis~\cite{wu2016merlin,li2018emphasis}, and to end-to-end neural based synthesis~\cite{shen2018natural,ping2018clarinet,ren2019fastspeech,ren2020fastspeech}. Neural network based end-to-end TTS models usually first convert input text to acoustic features (e.g., mel-spectrograms) and then transform mel-spectrograms into audio samples with a vocoder. Griffin-Lim~\cite{griffin1984signal} is a popular vocoder to reconstruct voices given linear-spectrograms. Other neural vocoders such as WaveNet~\cite{oord2016wavenet} and WaveRNN~\cite{kalchbrenner2018efficient} directly generate waveform conditioned on acoustic features. SVS systems are mostly inspired by TTS and follow the basic components in TTS such as text-to-audio alignment, parametric acoustic modeling and vocoder.

\paragraph{Singing Voice Synthesis}
Previous works have conducted studies on SVS from different aspects, including lyrics-to-singing alignment~\cite{fujihara2011lyricsynchronizer,chien2016alignment,gupta2018semi}, parametric synthesis~\cite{kim2018korean,blaauw2017neural}, acoustic modeling~\cite{nishimura2016singing,nakamura2019singing}, and adversarial synthesis~\cite{chandna2019wgansing,lee2019adversarially,hono2019singing}. \citet{blaauw2017neural} leverage the WaveNet architecture and separates the influence of pitch and timbre for parametric singing synthesis. \citet{lee2019adversarially} introduce adversarial training in linear-spectrogram generation for better voice quality. Jukebox~\cite{dhariwal2020jukebox} conditions on artist and lyrics to generate lots of musical styles and realistic singing voice with accompaniments together, while our work focuses on generating singing voice after vocal separation. Previous SVS systems usually obtain lyrics-to-singing alignment by human labeling or through the combination of human labeling and additional alignment tools, while DeepSinger builds SVS system from scratch (only original song-level lyrics and audio without any alignment information). Furthermore, most previous SVS systems use complicated acoustic parameters while DeepSinger generates linear-spectrograms directly based on a simple feed-forward Transformer model. 

\paragraph{Text-to-Audio Alignment}
Text-to-audio alignment is a key component for both TTS and SVS, which aims to obtain the duration of the pronunciation or the singing for each word/character/phoneme. TTS usually leverages a hidden Markov model (HMM) speech recognizer for phoneme-audio alignment~\cite{mcauliffe2017montreal}. For lyrics-to-singing alignment in SVS, traditional methods usually leverage the timing information from musical structure such as chords and chorus, or directly use musical score to align lyrics. However, such methods either need the presence of background accompaniments or require professional singing recordings where the notes are correctly sung, which are not practical for SVS. Recent works also leverage the alignment methods used in TTS for lyrics-to-singing alignment. \citet{gupta2018semi} propose a semi-supervised method for lyrics and singing alignment using transcripts from an automatic speech recognition system. \citet{sharma2019automatic,gupta2019acoustic} propose to align polyphonic music by adapting a solo-singing alignment model. \citet{lee2019adversarially} leverage laborious human labeling combined with additional tools for alignment. The works on lyrics-to-singing alignment either leverage a large amount of speaking voices such as LibriSpeech for pre-training or need human labeling efforts. In this work, we propose an automatic speech recognition based alignment model that is based on an encoder-attention-decoder framework to align lyrics and audio first in song level and then in sentence level, which does not need additional data or human labeling.

\paragraph{Training Data Mined From the Web}
A variety of tasks collect training data from the Web, such as the large-scale web-crawled text dataset ClueWeb~\citep{callan2009clueweb09} and Common Crawl\footnote{\url{https://commoncrawl.org/}} for language modeling~\citep{yang2019xlnet}, LETOR~\citep{qin2010letor} for search ranking~\citep{cao2007learning}, and WebVision~\citep{li2017webvision} for image classification. Similar to these works, collecting singing data from music websites also needs a lot of specific data processing, data mining, and data modeling techniques, including voice separation, alignment modeling, data filtration, and singing modeling.

\section{DeepSinger}

In this section, we introduce DeepSinger, a multi-lingual multi-singer singing voice synthesis (SVS) system, which is built from scratch by leveraging singing training data mined from music websites. We first describe the pipeline of DeepSinger and briefly introduce each step in the pipeline, and then introduce the formulation of the lyrics-to-singing alignment model and singing model. 

\subsection{Pipeline Overview}
\label{sec_pipline}
In order to build an SVS system from scratch by leveraging singing training data mined from music websites, as shown in Figure~\ref{arch_pipeline}, DeepSinger consists of several data processing, mining, and modeling steps, including 1) data crawling, 2) singing and accompaniment separation, 3) lyrics-to-singing alignment, 4) data filtration, and 5) singing modeling. Next, we introduce each step in details.

\paragraph{Data Crawling} 
In order to obtain a large amount of songs from the Internet, we crawl tens of thousands of songs and their lyrics from a well-known music website. Our crawled songs cover three languages (Chinese, Cantonese and English) and hundreds of singers. We perform some rough filtration and cleanings on the crawled dataset according to song duration and data quality: 
\begin{itemize}[leftmargin=*]
    \item We filter the songs which are too long (more than 5 minutes) or too short (less than 1 minute).
    \item We filter the songs in concert version based on their names since they are usually noisy.
    \item We remove some songs performed by music bands or multiple singers. 
    \item We also perform data cleaning on the crawled lyrics by removing some meta information (such as composer, singer, etc.) and unvoiced symbols.
    \item We keep the separation marks between each sentence in the lyrics for further sentence-level alignment and segmentation.
\end{itemize}
We convert the lyrics into phoneme sequence using open source tools: 1) we use phonemizer\footnote{\url{https://github.com/bootphon/phonemizer}} to convert English and Cantonese lyrics into corresponding phonemes, and 2) use pypinyin\footnote{\url{https://github.com/mozillazg/python-pinyin}} to convert Chinese lyrics into phonemes.

\paragraph{Singing and Accompaniment Separation}
Since the singing voices are mixed with the accompaniments in almost all the crawled songs, we leverage Spleeter~\cite{hennequin2019spleeter}, an open-source music source separation tool that achieves state-of-the-art accuracy, to separate singing voices from accompaniments and extract singing voices from the crawled songs. We normalize the loudness of the singing voices to a fixed loudness level (-16 LUFS\footnote{LUFS is short for loudness units relative to full scale.} with the measure of ITU-R BS.1770-3\footnote{ITU-R BS.1770-3 is a standard to measure audio loudness level. You can refer to \url{https://www.itu.int/rec/R-REC-BS.1770/en}.}).

\paragraph{Lyrics-to-Singing Alignment}
The duration of each phoneme determines how long each phoneme can be sung, and is critical for an SVS system to learn alignments between phonemes and acoustic features automatically. Therefore, we design a lyrics-to-singing alignment model to extract the duration of each phoneme first in sentence level and then in phoneme level. We describe the details of the alignment model in Section~\ref{sec_alignment}.

\paragraph{Data Filtration}
After the lyrics-to-singing alignment, we found there are some misaligned lyrics and singing voices, which may be due to the following reasons: 1) the voice quality of some songs is poor, since they are not recorded in a professional recording studio or contain mixed chorus; 2) some lyrics are not belong to their songs, due to errors from music websites; 3) the quality of some separated singing voices is poor since the accompaniments cannot be totally removed. Therefore, we filter those misaligned singing voices according to their alignment quality, where we use the splitting reward introduced in Section~\ref{sec_alignment}, as the filtration criterion and filter singing data with the splitting reward lower than a threshold.

\paragraph{Singing Modeling}
After the previous steps of data crawling, separation, alignment, and filtration, the mined singing data are ready for singing modeling. We design a FastSpeech based singing model that takes lyrics, duration, pitch information, as well as a reference audio as input to generate the singing voices. We introduce the details of the singing model in Section~\ref{sec_singing_model}.

\subsection{Lyrics-to-Singing Alignment}
\label{sec_alignment}
Lyrics-to-singing alignment is important for SVS to decide how long each phoneme is sung in synthesized voices. Previous works~\cite{lee2019adversarially,yi2019singing} usually leverage human labeling to split songs into sentences and then conduct phoneme alignment within each sentence by leveraging an HMM (hidden markov model) based speech recognition model. In this paper, we propose a new alignment model to extract the duration of each phoneme, by leveraging raw lyrics and song recordings, without relying on any human labeling efforts.

It is hard to directly extract the duration of each phoneme given lyrics and audio from a whole song. Instead, we use a two-stage alignment pipeline that first splits a whole song into sentences, and then extracts the phoneme duration in each sentence, as described as follows:
\begin{itemize}[leftmargin=*]
    \item We train our alignment model (which is introduced in the next paragraphs) with the lyrics and audio in a whole song, and use it to split the whole song into aligned lyrics and audio in sentence level. We segment an audio by the frames that are aligned to the separation marks in raw lyrics.
    \item We continue to train the previous alignment model with the aligned lyrics and audio in the sentence level that are obtained in the previous step. We use this trained alignment model to obtain lyrics-to-singing alignments in the phoneme level. We further design a duration extraction algorithm to obtain the duration of each phoneme from the lyrics-to-singing alignments. 
\end{itemize}
Specifically, we train the alignment model based on automatic speech recognition with some strategies of greedy training and guided attention, and design a dynamic programming based duration extraction to obtain the phoneme duration. We introduce the two modules as follows.

\begin{figure}[!h] 
	\centering
		\centering
		\includegraphics[width=0.5\textwidth]{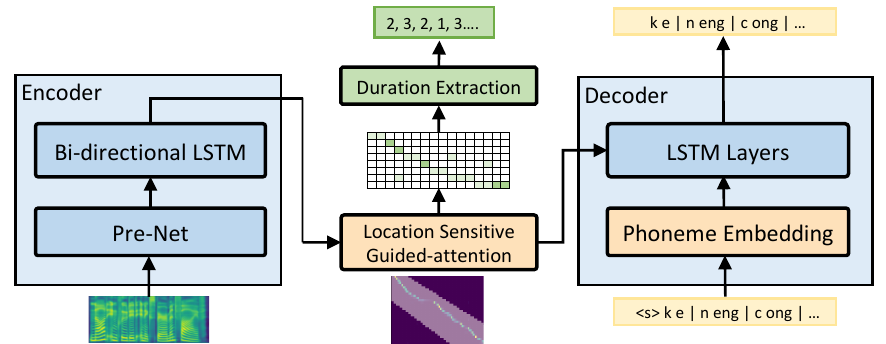}
		\caption{The alignment model based on the architecture of automatic speech recognition.}
		\label{arch_alignmodel}
		\vspace{-0.5cm}
\end{figure}

\begin{figure}[!h] 
	\centering
	\includegraphics[width=0.45\textwidth,trim={0.3cm 0.0cm 0cm 0.0cm}, clip=true]{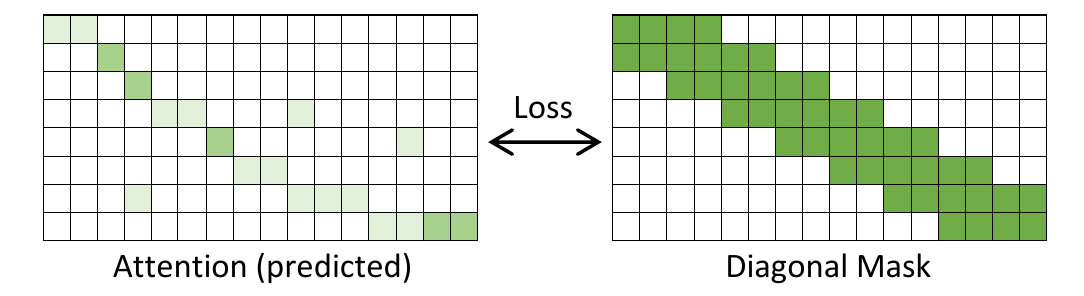}
	\caption{The details of the guided attention.}
	\label{attn_kd}
	\vspace{-0.4cm}
\end{figure}

\paragraph{Alignment Model Training} 
We leverage an automatic speech recognition model with location sensitive attention~\cite{shen2018natural} to obtain lyrics-to-singing alignments, as shown in Figure~\ref{arch_alignmodel}. The alignment model consists of a bi-directional LSTM encoder and an LSTM decoder, which takes singing audio (mel-spectrograms) as input and lyrics (phoneme) as output. The attentions between the encoder and decoder are regarded as the alignments between audio and lyrics. To ensure alignment accuracy, we propose two techniques to help the training of the alignment model:  
\begin{itemize}[leftmargin=*]
\item Greedy Training. It is hard to train the model by taking the whole song as input. Instead, we train the alignment model with a greedy strategy: 1) First, only a small part of the whole audio and lyrics from the beginning of each song are fed into the model for training. 2) Second, the length of the audio and lyrics are gradually increased for training. 3) Finally, the whole song are fed into the model for training. By increasing the difficulty of this training task, the model can learn the ability to align in the very beginning and gradually to lean to align the whole song. 
\item Guided Attention. In order to further help the model learn reasonable alignments, we leverage guided attention~\cite{tachibana2018efficiently} to constrain the attention alignments. Guided attention is based on prior knowledge that the attentions between lyrics and audio should be diagonal and monotonic. We first construct a diagonal mask matrix $M\in \{0, 1\}^{T \times S}$ (shown in Figure~\ref{attn_kd}) as follows:
$$
M_{i,j}=\left\{
\begin{array}{rcl}
1 & & {i* \frac{S}{T}-W \leq j \leq i*\frac{S}{T}+W}\\
0 & & {else}\\
\end{array} \right. ,
$$
where $W$ is the bandwidth of $M$ representing how many elements equal to one, $S$ is the length of the mel-spectrogram frames and $T$ is the length of the phoneme sequence. The guided attention loss is defined as $-M*A$ where $A$ is the weights of the encoder-decoder attention in the alignment model, and the guided attention loss will be added to the original loss of the alignment model. If the attention weight $A$ is far from diagonal, it will be pushed towards diagonal for better alignment quality.
\end{itemize}

\paragraph{Duration Extraction}
After we get the lyrics-to-singing alignments from the alignment model, an intuitive way to extract the phoneme duration is to count how many consecutive frames of mel-spectrograms that a phoneme is attending. However, it is non-trivial considering the following reasons: 1) the attention map from the alignment model is not always diagonal and monotonic; 2) a mel-spectrogram frame can be attended by multiple phonemes and it is hard to decide which phoneme actually corresponds to this mel-spectrogram frame. In order to extract the duration from the attention map accurately, we design a novel duration extraction algorithm to compute the duration $D \in \mathbb{R}^T$ for each phoneme in a sequence, which satisfies:
\begin{equation}
\begin{aligned}
\sum_{i=1}^{T}{[D]_{i}} = S,
\label{eqn_xx1}
\end{aligned}
\end{equation}
where $T$ is the length of a phoneme sequence and $S$ is the length of mel-spectrogram frames. The extraction of $D$ consists of several steps:
\begin{itemize}[leftmargin=*]
\item First, we obtain the attention alignment $A \in \mathbb{R}^{T \times S}$ from the alignment model. The $i$-th row ${A_{i}} \in \mathbb{R}^S, i \in [1, T]$ is a non-negative probability distribution, where each element $A_{i,j}$ represents the probability of the $i^{th}$ phoneme attending to the $j^{th}$ mel-spectrogram frame. 
\item Second, we define the splitting boundary vector $B \in R^{T+1}$, which splits the mel-spectrogram frames into $T$ segments corresponding to the $T$ phonemes. The splitting reward $\mathcal{O}$ for some certain split points is defined as follows:
\begin{equation}
\begin{aligned}
\mathcal{O} = \sum_{i=1}^{T}{ \sum_{j=B_{i}}^{B_{i+1}-1}{A_{i,j}}}.
\label{eqn_xx2}
\end{aligned}
\end{equation}
\item Third, the goal of duration extraction becomes to find the best split boundary vector $B$ by maximizing the splitting reward $\mathcal{O}$, and then to extract the duration according to the split boundary vector. Maximizing the splitting reward can be solved by a standard dynamic programming (DP) algorithm. The details of the DP algorithm for duration extraction is summarized in Section \ref{sec:tab_appendix_alignment}. 
\item Forth, after we get the best splitting boundary vector $B \in R^{T+1}$ based on the DP algorithm, the duration $D$ can be simply calculated by 
\begin{equation}
\begin{aligned}
D_{i} = B_{i+1}-B_{i}.
\label{eqn_xx3}
\end{aligned}
\end{equation}

\end{itemize}

\subsection{Singing Modeling}
\label{sec_singing_model}
\begin{figure}[!th] 
	\centering
		\includegraphics[width=0.48\textwidth]{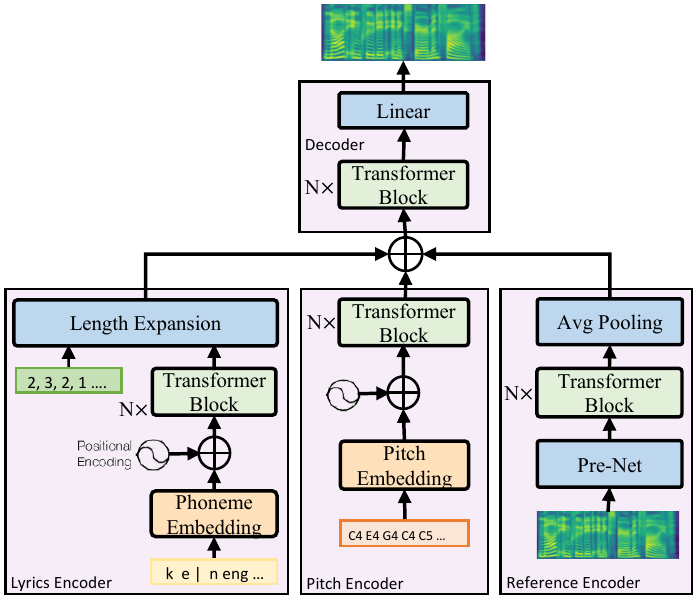}
	\caption{The architecture of the singing model.}
	\label{arch_singing_model}
	\vspace{-0.3cm}
\end{figure}

After we get the duration of each phoneme in lyrics with the alignment model, in this section, we build our singing model to synthesize singing voices. Our singing model is based on FastSpeech~\citep{ren2019fastspeech,ren2020fastspeech}, which is a feed-forward Transformer network to synthesize speech in parallel. FastSpeech leverages a length regulator to expand the phoneme sequence to the length of the target speech according to the duration of each phoneme, which is suitable for singing modeling since duration information is needed by default in SVS. Different from FastSpeech, our singing model 1) leverages separate lyrics and pitch encoders for pronunciation and pitch modeling, 2) designs a reference encoder to capture the timbre of a singer instead of a simple singer embedding, to improve the robustness of the learning from noisy data, and 3) directly generates linear-spectrograms instead of mel-spectrograms and using Griffin-Lim~\citep{griffin1984signal} for voice synthesis. As shown in Figure~\ref{arch_singing_model}, we describe each module in our singing model as follows.

\paragraph{Lyrics Encoder} The lyrics encoder consists of 1) a phoneme embedding lookup table to convert the phoneme ID into embedding vector, 2) several Transformer blocks~\cite{vaswani2017attention} to convert the phoneme embedding sequence into a hidden sequence, and 3) a length expansion operation to expand the length of the hidden sequence to match the length of the target linear-spectrograms according to the phoneme duration.

\paragraph{Pitch Encoder} The pitch encoder consists of 1) a pitch embedding lookup table to convert the pitch ID into embedding vector, and 2) several Transformer Blocks to generate pitch hidden sequence. Since the pitch sequence is directly extracted from the audio in the training set and has the same length with the linear-spectrogram sequence, we do not need any length expansion as used in the lyrics encoder.  
\paragraph{Reference Encoder} The reference encoder consists of 1) a pre-net to preprocess the linear-spectrograms of the reference audio, 2) several Transformer blocks to generate a hidden sequence, and 3) an average pooling on the time dimension to compress the hidden sequence into a vector, which is called as the reference embedding and contains the timbre information of the speaker. The reference embedding is broadcasted along the time dimension when adding with the outputs of the lyrics and pitch encoder. The reference encoder shows advantages over singer embedding especially when the singing training data are noisy:
\begin{itemize}[leftmargin=*]
    \item A singer embedding will average the timbre as well as voice characteristics of the singing data with different noise levels. Therefore, the singer embedding contains the characteristics of noise audio, which makes the synthesized voice noisy.
    \item A reference encoder only learns the characteristics from a reference audio, which can ensure the model can synthesize clean voice given clean reference audio.
\end{itemize}
\paragraph{Decoder} The decoder consists of 1) several Transformer blocks to convert the frame-wise addition of the outputs of the three encoders into a hidden sequence, and 2) a linear layer to convert the hidden sequence into linear-spectrograms. Finally, we use Griffin-Lim~\citep{griffin1984signal} to directly synthesize singing voices given the predicted linear-spectrograms.

\begin{figure}[!h] 
	\centering
	\includegraphics[width=0.48\textwidth,trim={0cm 0cm 0cm 0cm}, clip=true]{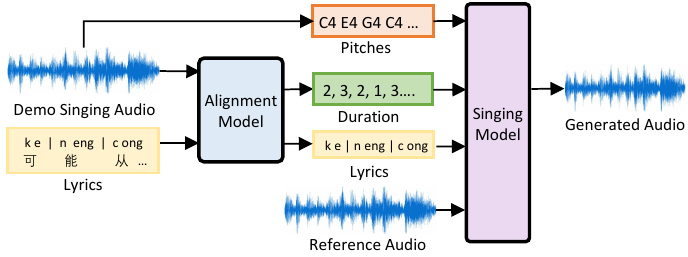}
	\caption{The inference process of singing voice synthesis.}
	\label{fig_infer_demosinging}
	\vspace{-0.3cm}
\end{figure}

The training and inference processes of the singing model are described as follows.
\begin{itemize}[leftmargin=*]
    \item Training. We extract the pitches from the training audio and transform the audio into linear-spectrograms. Then we feed the phoneme sequence together with duration, pitch sequence, and linear-spectrograms into the lyrics encoder, pitch encoder, and reference encoder correspondingly. We minimize the mean square error (MSE) loss between the output linear-spectrograms and the reference linear-spectrograms to optimize the model. 
    \item Inference. DeepSinger synthesizes voices from lyrics and demo singing audio, as shown in Figure~\ref{fig_infer_demosinging}. We extract the duration and pitch information similar to that used in the training. The reference audio can be any singing or speaking audios to provide the voice characteristics such as timbre for the synthesized voices. 
\end{itemize}

\section{The Mined Singing Dataset}
\label{sec_singing_wild}
In this section, we briefly introduce our mined singing dataset (named as Singing-Wild, which means singing voice dataset in the wild). 

Singing-Wild contains more than 90 hours singing audio files in multiple languages (Chinese, Cantonese and English), and multiple singers, and the corresponding lyrics, phonemes, and singing-phoneme alignment files. In addition, we provide some meta data which store the song name, singer information and  downloading links. The detailed data statistics of the Singing-Wild dataset are listed in Table~\ref{tab_dataset_stats}. We further analyze some statistics in details, including the duration distributions of the singing voices in the sentence level and phoneme level, and the pitch distributions.

\begin{table}[!h]
\small
	\centering
	\begin{tabular}{ l | c | c | c | c }
	\toprule
		Language &  \#Singers  & \#Songs & \#Sentences & Duration (hours)  \\
		\midrule
		\textit{Chinese} & 25 & 1910 & 52109 & 62.61  \\
		\textit{Cantonese} & 41 & 322 & 14853 & 18.83 \\
		\textit{English} & 23 & 741 & 13217 & 11.44  \\
		\bottomrule
		\textit{Total} & 89 & 2973 & 70179 & 92.88  \\
		\bottomrule
	\end{tabular}
	\caption{The statistics of the Singing-Wild dataset.}
	\label{tab_dataset_stats}
	\vspace{-0.7cm}
\end{table}

\paragraph{Sentence-Level Duration Distribution}
We segment the songs into sentence-level singing voices using the lyrics-to-singing alignment model mentioned in Section \ref{sec_alignment}. We plot the sentence-level duration distribution of the singing voices in Table \ref{fig_sent_duration} in Section \ref{sec:sw_detail}. We find that the distributions of Chinese and Cantonese are very similar, and most of sentences in these languages are around 4 seconds, while most sentences in English are around 2.5 seconds. 

\paragraph{Phoneme-Level Duration Distribution}
We align the phoneme sequence to the singing audio frames using the lyrics-to-singing alignment model mentioned in Section \ref{sec_alignment}. We plot the distributions of the phoneme duration in Table \ref{fig_ph_duration} in Section \ref{sec:sw_detail}. We find that most of the phoneme duration on English is distributed around 10 ms to 100 ms, while the individual phoneme duration on Chinese and Cantonese is longer than English, which is distributed around 10 ms to 200 ms.

\paragraph{Pitch Distribution}
We extract the F0 (fundamental frequency, also known as pitch) from the audio using Parselmouth\footnote{\url{https://github.com/YannickJadoul/Parselmouth}} and convert F0 to a sequence of international pitch notation\footnote{\url{http://www.flutopedia.com/octave\_notation.htm}}. The international pitch notation expresses a note with a musical note name and a number to identify the octave of a pitch, e.g., C4, D4 and A5. We also apply the median filter to the note sequence to remove noises. We plot the pitch distributions in Table \ref{fig_pitch_dist} in Section \ref{sec:sw_detail}. We can see that most of the pitches are distributed around C2 to C4 on all languages.

\section{Experiments and Results}
In this section, we first describe the experimental setup, report the accuracy of the alignment model in DeepSinger, and then evaluate the synthesized voices both quantitatively in terms of pitch accuracy and qualitatively in terms of mean opinion score (MOS). Finally, we conduct some analyses of DeepSinger.

\subsection{Experimental Setup}
\subsubsection{Datasets} We use our mined Singing-Wild dataset to train the multi-lingual and multi-singer singing model. For each language, we randomly select 5 songs of different singers and pick 10 sentences in each song (totally 50 sentences) to construct the test set, and similarly construct another 50 sentences as the valid set, and use the remaining songs as the training set. Inspired by~\citet{zhang2019learning}, we also leverage extra multi-speaker TTS datasets to help the training of the singing model and improve the quality of the generated voices. We use 1) THCHS-30~\cite{THCHS30_2015} dataset for Chinese, which consists of 10893 sentences (27 hours) from 30 speakers, 2) a subset of Libritts~\cite{zen2019libritts} for English, which consists of 7205 sentences (12 hours) from 32 speakers, and 3) an internal dataset for Cantonese, which consists of 10000 sentences (25 hours) from 30 speakers. 

For the audio data, we first convert the sampling rate of all audios to $22050$Hz, and then convert the audio waveform into mel-spectrograms for alignment model and linear-spectrograms for singing model. The frame size and hop size of both the mel-spectrograms and linear-spectrograms are set to 1024 and 256 respectively. For the text data, we convert text sequence into phoneme sequence with the open-source tools as mentioned in Section~\ref{sec_pipline}.

\subsubsection{Alignment Model}

\paragraph{Data Preprocessing} We first mark the non-vocal frames whose pitches cannot be successfully extracted or volumes are lower than a threshold. Then we detect the non-vocal segments which contain at least 10 consecutive non-vocal frames. We replace each non-vocal segment with 10 silence frames, which can significantly shorten the audio and make the alignment training easier.

\paragraph{Model Configuration} As shown in Figure~\ref{arch_alignmodel}, the alignment model consists of a mel-spectrogram encoder and a phoneme decoder. The pre-net in the encoder consists of a 3-layer 1D convolution with hidden size and kernel size of 512 and 9 respectively. The hidden size of the bi-directional LSTM in the encoder is 512. The decoder consists of a phoneme embedding lookup table with dimension of 512, and a 2-layer LSTM with hidden size of 1024.

\paragraph{Model Training} The alignment model is trained with a batch size of 64 sentences on one NVIDIA V100 GPU. We use Adam optimizer~\cite{kingma2014adam} with $\beta_1 = 0.9$, $\beta_2 = 0.999 $, $\epsilon = 10^{-6} $ and a learning rate of $10^{-3}$ that is exponentially decaying to $10^{-5}$ after 30,000 iterations. We also apply L2 regularization with weight of $10^{-6}$. We use greedy training to better learn the alignment model. Specifically, we train the alignment model with the first 10\% length of lyrics and audio in each song, and then we gradually increase the length by 2\% every epoch. It takes 100 epochs for training till convergence. We fine-tune the alignment model with the aligned lyrics and audio in the sentence level to further obtain the phoneme-level alignment.

When training the alignment model with the lyrics and audio in the whole song level, it is difficult to store all the necessary activations in GPU memory for back-propagation due to the extremely long sequence. Therefore, we apply the truncated back-propagation through time (TBTT)~\cite{jaeger2002tutorial}, which is widely used in training RNN models. TBTT does not need for a complete backtrack through the whole sequence and thus saves GPU memory. We detail the whole training procedure in Section \ref{sec:tab_appendix_alignment}.

\subsubsection{Singing Model}
\paragraph{Model Configuration} Our singing model consists of 4, 2, 4 and 4 Transformer blocks in lyrics encoder, pitch encoder, reference encoder and decoder respectively and the output linear layer converts the 384-dimensional hidden into 513-dimensional linear-spectrograms. Other configurations follow ~\citet{ren2019fastspeech} and we list them in Section \ref{sec:append_singing}.

\paragraph{Model Training and Inference} We train the singing model with a batch size of totally 64 sequences on 4 NVIDIA V100 GPUs. We use Adam optimizer with $\beta_{1}= 0.9$, $\beta_{2} = 0.98$, $\varepsilon = 10^{-9}$ and follow the same learning rate schedule in \citet{vaswani2017attention}. It takes 160k steps for training till convergence. For each singer in the test set, we choose a high-quality audio of this singer from the training set as the reference audio to synthesize voices for evaluation.

\subsection{Accuracy of Alignment Model}
We use two metrics to measure the alignment accuracy: 1) percentage of correct segments (PCS)~\cite{dzhambazov2017knowledge} for sentence-level alignment, which measures the ratio between the length of the correctly aligned segments and the total length of the song, 2) average absolute error (ASE)~\cite{mesaros2008automatic} for phoneme-level alignment, which measures the phoneme boundary deviation. 

For the 50 test sentences in each language, we manually annotate the sentence-level and phoneme-level boundaries for the three languages as the ground truth to evaluate the accuracy of the alignment model. It is easy for human to annotate the sentence-level boundaries, while hard to annotate the phoneme-level boundaries. Therefore, we annotate the word/character-level (word level for English and character level for Chinese and Cantonese) boundaries instead of phoneme-level boundaries. 

The PCS and ASE for the three languages are shown in Table~\ref{result_ase}. It can be seen that PCS on all the three languages are above 80\%, while ASE are below 100ms, which demonstrates the high accuracy of our alignment model in both sentence level and word/character level, considering the distributions of sentence-level and phoneme-level duration shown in Figure \ref{fig_sent_duration} and \ref{fig_ph_duration}.

We compare the accuracy of our alignment model on Chinese with Montreal Forced Aligner (MFA)~\cite{mcauliffe2017montreal}, an open-source system for speech-text alignment with good performance, which is also trained without any manually alignment annotations. Since MFA cannot support the alignments for long sentence (above 60 seconds) well, we only compare our model with MFA in term of character-level ASE. MFA achieves 78.5ms character-level ASE in the test set while our model achieves 76.3ms, with very close accuracy. However, our model supports the alignments for whole song while MFA cannot, which demonstrates the advantages of our alignment model. 

\begin{table}[!h]
\centering
  \begin{tabular}{ l   c   c  }
    \toprule
    Language & PCS (sentence-level) & ASE \\
    \midrule
	    \textit{Chinese} & 83.1\% & 76.3 ms   \\
		\textit{Cantonese} & 81.2\% & 85.2 ms  \\
		\textit{English} & 77.4\% & 92.5 ms  \\
    \bottomrule
  \end{tabular}
  \caption{The accuracy of the alignment model on the three languages, in terms of the sentence-level metric: percentage of correct segments (PCS) and word/character-level metric: average absolute error (ASE). For ASE, we use word level for English and character level for Chinese and Cantonese.}
  \label{result_ase}
	\vspace{-0.7cm}
\end{table}

\subsection{Voice Quality}
We evaluate the quality of the synthesized voices by our singing model, both quantitatively with pitch accuracy and qualitatively with mean opinion score (MOS).

\paragraph{Quantitative Evaluation}
We evaluate the accuracy of the pitches in the synthesized singing voices, following~\citet{lee2019adversarially}. We first extract the fundamental frequency (F0) sequence from the synthesized audio using Praat~\cite{boersma2002praat}, and then convert it into pitch sequence. We calculate the frame-wise accuracy of the extracted pitch sequence with regard to the ground-truth pitch sequence that the singing model conditions on. We also provide the upper bound of the pitch accuracy by calculating the pitch accuracy of the voices reconstructed from the ground-truth linear-spectrograms, with regard to the ground-truth voices. As shown in Table~\ref{tab_pitch_results}, the upper bounds of the pitch accuracy on the three languages are higher than 95\%, while DeepSinger can generate songs with pitch accuracy higher than 85\% on three languages. Considering that human singing to a music score can usually achieve about 80\% accuracy according to~\citet{lee2019adversarially}, the accuracy achieved by DeepSinger is high enough.

\begin{table}[!h]
	\centering
	\begin{tabular}{ l | c | c | c}
		\toprule
		     Setting    &  \textit{Chinese} & \textit{Cantonese} & \textit{English} \\
		\midrule
		     DeepSinger       &  87.60\%      & 85.32\%  &  86.02\% \\
		     Upper Bound           &  96.28\%      & 95.62\%  &  96.06\% \\
		\bottomrule
	\end{tabular}
	\vspace{0.1cm}
	\caption{The pitch accuracy of DeepSinger and the corresponding upper bound.}
	\label{tab_pitch_results}
	\vspace{-0.6cm}
\end{table}

\paragraph{Qualitative Evaluation}
We conduct MOS (mean opinion score) evaluation on the test set to measure the quality of the synthesized voices, following \citet{blaauw2017neural}. Each audio is listened by at least 20 testers, who are all native speakers for each language. We compare the MOS of the synthesized audio samples among the following systems: 1) \textit{GT}, the ground-truth audio; 2) \textit{GT (Linear+GL)}, where we synthesize voices based on the ground-truth linear-spectrograms using Griffin-Lim; 3) \textit{DeepSinger}, where the audio is generated by DeepSinger. The results are shown in Table~\ref{tab_mos_results}. It can be seen that DeepSinger synthesizes high-quality singing voices, with only 0.34, 0.76 and 0.43 MOS gap to the \textit{GT (Linear+GL)} upper bound on the three languages. Our audio samples are shown in the demo website\footnote{\url{https://speechresearch.github.io/deepsinger/}}.

\begin{table}[!h]
	\centering
	\begin{tabular}{ l | c | c | c  }
		\toprule
		Setting                  &  Chinese       & Cantonese      & English        \\
		\midrule
		\textit{GT}             & 4.36 $\pm$ 0.08  & 4.38 $\pm$ 0.09  & 4.15 $\pm$ 0.10   \\
		\textit{GT (Linear+GL)} & 4.12 $\pm$ 0.07  & 4.18 $\pm$ 0.09  & 3.95 $\pm$ 0.10   \\
		\midrule
		\textit{DeepSinger}     & 3.78 $\pm$ 0.10  & 3.42 $\pm$ 0.10  & 3.52 $\pm$ 0.11   \\
		\bottomrule
	\end{tabular}
	\caption{The MOS of DeepSinger with 95\% confidence intervals on the three languages.}
	\label{tab_mos_results}
	\vspace{-0.8cm}
\end{table}

\begin{table}[!h]
	\centering
	\begin{tabular}{ l | c  }
		\toprule
		Setting &  MOS   \\
		\midrule
		\textit{DeepSinger}                          & 3.78 $\pm$ 0.10  \\
	    \midrule
	    \textit{DeepSinger w/o reference encoder}      & 3.36 $\pm$ 0.11  \\
		\textit{DeepSinger w/o TTS data}               & 3.25 $\pm$ 0.12  \\
		\textit{DeepSinger w/o multilingual}           & 3.79 $\pm$ 0.08  \\
		\bottomrule
	\end{tabular}
	\caption{The MOS with 95\% confidence intervals for different methods.}
	\label{tab_mos_ablation}
	\vspace{-0.8cm}
\end{table}

\subsection{Method Analyses}
We conduct experimental studies on Chinese to analyze some specific designs in DeepSinger, including the effectiveness of the reference encoder, the benefits of leveraging TTS data for auxiliary training, and the influence of multilingual training on voice quality. We introduce the analyses as follows.

\paragraph{Reference Encoder}
We analyze the effectiveness of the reference encoder in DeepSinger to handle noisy training data, from three perspectives:
\begin{itemize}[leftmargin=*]
    \item We compare DeepSinger (denoted as \textit{DeepSinger} in Table~\ref{tab_mos_ablation}) to the system without reference encoder but instead with a singer embedding to differentiate multiple singers (denoted as \textit{DeepSinger w/o reference encoder} in Table~\ref{tab_mos_ablation}). It can be seen that \textit{DeepSinger} generates singing voices with higher quality than \textit{DeepSinger w/o reference encoder}, which demonstrate the advantages of reference encoder than singer embedding. 
    \item We further analyze how the reference encoder takes effect with reference audios in different noise levels. As shown in Table~\ref{tab_mos_diff_ref}, we choose clean, normal, and noisy reference audios and evaluate the MOS of the synthesized speech. According to the MOS, it can be seen that clean voice can be synthesized given clean reference audio while noisy reference leads to noisy synthesized voice, which indicates that the reference encoder can learn the characteristics from the reference audio, verifying the analyses in Section~\ref{sec_singing_model}.
    \item We also compare \textit{DeepSinger w/o reference encoder} with \textit{DeepSinger} in term of preference scores\footnote{We introduce the preference score in Section \ref{sec:appendix_eval_metrics}.}. We evaluate the preference scores on the singing training data with two noisy levels: clean singing data and noisy singing data. As shown in Table \ref{tab:preference_se_re}, \textit{DeepSinger} largely outperforms \textit{DeepSinger w/o reference encoder} on noisy singing training data while is slightly better than \textit{DeepSinger w/o reference encoder} on clean singing training data. The results indicate that 1) singer embedding averages the timbre of the training data with different noise levels and thus performs differently when training with singing data in different noise levels; 2) our singing model can be tolerant of the noisy data with the help of reference encoder and can still generate good voice for noisy training data so long as the clean reference audio is given.
    
\end{itemize}

\begin{table}[!h]
	\centering
	\begin{tabular}{ l | c | c }
		\toprule
				Setting &  Ref MOS & Syn MOS \\
		\midrule
		\textit{DeepSinger (Clean Ref)}   & 4.38 $\pm$ 0.09   & 3.78 $\pm$ 0.10  \\
		\textit{DeepSinger (Normal Ref)}  & 4.12 $\pm$ 0.10   & 3.42 $\pm$ 0.12   \\
		\textit{DeepSinger (Noisy Ref)}   & 3.89 $\pm$ 0.10   & 3.27 $\pm$ 0.11   \\
		\bottomrule
	\end{tabular}
	\caption{The MOS with 95\% confidence intervals. We generate singing with different (clean, normal and noisy) reference audios of the same singer. Ref and Syn MOS represent the MOS of the reference audio and synthesized audio respectively.}
	\label{tab_mos_diff_ref}
	\vspace{-0.8cm}
\end{table}

\begin{table}[!h]
	\centering
	\begin{tabular}{ c | c | c | c }
	\toprule
	 Data Setting& \textit{DeepSinger} & Neutral & \textit{DeepSinger w/o RE}     \\
	 \midrule
	Clean Singing &  28.50\%  & 45.30\%  & 26.20\%   \\
    Noisy Singing &     66.20\%  & 26.80\%  & 7.00\%   \\
	\bottomrule 
	\end{tabular}
	\caption{The preference scores of singing model with reference encoder (our model, denoted as \textit{DeepSinger}) and singing model with speaker embedding (denoted as \textit{DeepSinger w/o RE}) when training with clean and noisy singing data.}
	\label{tab:preference_se_re}
	\vspace{-0.5cm}
\end{table}

\paragraph{TTS Training Data}
We further conduct experiments to explore the effectiveness of extra multi-speaker TTS data for auxiliary training. We train the singing model without multi-speaker TTS data (denoted as \textit{DeepSinger w/o TTS data}) and compare it to DeepSinger. As shown in Table~\ref{tab_mos_ablation}, the voice quality of \textit{DeepSinger w/o TTS data} drops compared to that of \textit{DeepSinger}, indicating extra TTS data is helpful for the singing model, which is consistent with the previous work \cite{zhang2019learning}. To demonstrate that our mined Singing-Wild dataset is critical for singing model training, we train another singing model with only TTS data (denoted as \textit{DeepSinger (only TTS)}) and compare it to \textit{DeepSinger} in term of preference scores. The results are shown in Table~\ref{tab_mos_ablation_on_trainset}. It can be seen that DeepSinger largely outperforms the model trained with only TTS data, which demonstrates the importance of Singing-Wild dataset for SVS.

\begin{table}[!h]
	\centering
	\begin{tabular}{ l | c | c | c }
		\toprule
		     & \textit{DeepSinger}  & Neutral & \textit{DeepSinger (only TTS)}    \\
		\midrule
		Preference Score  & 65.00 \%  & 15.50 \%  & 19.50 \%   \\
		\bottomrule 
	\end{tabular}
	\caption{The preference scores of \textit{DeepSinger} and \textit{DeepSinger (only TTS)}. We choose one TTS audio as the reference audio to generate singing voices on Chinese singing test set.}
	\label{tab_mos_ablation_on_trainset}
	\vspace{-0.6cm}
\end{table}

\paragraph{Multilingual Training}
We then analyze whether multilingual training affects the quality of the synthesized voices for a certain language. We train a singing model with singing training data only in Chinese (denoted as \textit{DeepSinger w/o multilingual} in Table~\ref{tab_mos_ablation}) and compare it to DeepSinger. As can be seen, the voice quality of \textit{DeepSinger} is nearly the same as that of \textit{DeepSinger w/o multilingual}, which demonstrates that multilingual training does not affect the voice quality of the singing model. As a byproduct of multilingual training, DeepSinger can also perform cross-lingual singing voice synthesis, as described in Section~\ref{sec:appendix_xl}.

\section{Discussions}

While DeepSinger can synthesize reasonably good singing voices, expert studies show that the synthesized voices have unique styles and are very different from human voices:
\begin{itemize}
    \item DeepSinger does not contain breathing and breaks that are common in human singing voices, since AI has no physical constraints caused by human vocal organs.
    \item The synthesized singing voices do not have as rich and diverse expressiveness and emotion as human voices, because DeepSinger simply learns  average patterns from training data. 
\end{itemize}
    Those differences make it easy to distinguish synthesized voices from human singing voices. While one may think this is the disadvantage of DeepSinger, we would like to emphasize that our goal is not to clone human singing voices; in contrast, we target at generating beautiful AI singing voices with unique styles which can bring new artistic experiences to human.

\section*{Acknowledgments}

This work was supported in part by the National Key R\&D Program of China (Grant No.2018AAA0100603), Zhejiang Natural Science Foundation (LR19F020006), National Natural Science Foundation of China (Grant No.61836002), National Natural Science Foundation of China (Grant No.U1611461), National Natural Science Foundation of China (Grant No.61751209) and the Fundamental Research Funds for the Central Universities (2020QNA5024). This work was also partially funded by Microsoft Research Asia.

\bibliographystyle{ACM-Reference-Format}
\bibliography{deepsinger}

\clearpage
\appendix
\section{Reproducibility}\label{sec:reproducibility}

\subsection{Details in Data Crawling}
We use the python library Requests\footnote{https://github.com/psf/requests} to build the data crawler. We fetch the singer ids from the website first and then use the singer id S\_ID to visit the song page of the singer. Finally, we can download all available songs of each singer.

\subsection{Details in Lyrics-to-Singing Alignment}
\label{sec:tab_appendix_alignment}

\paragraph{Training Details}

\begin{algorithm}[h]
\caption{Alignment Model Training (with mini batch size 1)}\label{alg:align}
\begin{algorithmic}[1]
\State \textbf{Input}: Training dataset $\mathcal{D}=\{(M, Ph) \in (\mathcal{X}\times \mathcal{Y})\}$, where $(M, Ph)$ is a pair of mel-spectrograms and text, total training epoch $e$, TBTT chunk size $C$. 
\State \textbf{Initialize}: Set current maximum target length ratio $L$ = $10\%$, phoneme-to-mel position mapping $\mathcal{A}$, alignment model $\mathcal{M}$.

\For{each $i\in [0,e)$}
    \For{each $\textit{id}, (M, Ph) \in \mathcal{D}$}
        \State $ s = 0 , \textit{State}_\textit{dec} = \emptyset $
        \State set $T$ to the length of $Ph$, set $S$ to the length of $M$
        \While{ $ s < L * T $ }
            \State $ Ph_{chunk} = Ph[s:s+C] $
            \State $ M_{chunk} = M[\mathcal{A}[s]:\mathcal{A}[s]+ C*\frac{S}{T}] $

            \State  $\textit{loss}, \textit{State}_\textit{dec}, \mathcal{A}'$  =  $\mathcal{M}(\textit{State}_\textit{dec}, Ph_\textit{chunk}, M_\textit{chunk}) $
            \State Optimize $\mathcal{M}$ with \textit{loss}
            \State Update  $\mathcal{A}_\textit{id}[s:s+C]$ with $\mathcal{A}'$
            \State $ s = s + C $
        \EndWhile
    \EndFor
    \State $ L = L + 2\% $
\EndFor
\State \textbf{return} $\mathcal{M}, \mathcal{A}$

\end{algorithmic}
\end{algorithm}

The training of alignment model is shown as Algorithm~\ref{alg:align}\footnote{We show the training process when mini batch size is 1 for simplicity.}. In Line 3, we begin training the model for $e$ epochs. In Line 4, we traverse the dataset to fetch data id (\textit{id}), mel-spectrogram ($M$) and phone sequence ($Ph$). In Line 5, we use $s$ to represent the start position of phoneme subsequence $Ph_\textit{chunk}$ in the chunk training, and initialize the decoder states for TBTT training. In Line 9, we get the phoneme and mel-spectrogram subsequences ($Ph_\textit{chunk}$ and $M_\textit{chunk}$) for this chunk. In Line 10 to Line 12, we train the alignment model $\mathcal{M}$ and update the phoneme-to-mel position mapping $\mathcal{A}$ with $\mathcal{A}'$, where $\mathcal{A}'$ can be transformed from duration $D$ which is extracted by the duration extraction algorithm. When the training ends, we can get lyrics-to-singing alignment from $\mathcal{A}$. In our work, we set $C$ to $40$, and $\mathcal{T}$ to $100$.

\paragraph{Algorithm Details in Duration Extraction}

\begin{algorithm}[!h]
\caption{DP for Duration Extraction}\label{alg}
\begin{algorithmic}[1]
\State \textbf{Input}: Alignment matrix $A\in \mathbb{R}^{\mathcal{T} \times \mathcal{S}}$
\State \textbf{Output}: Phoneme duration $D \in \mathbb{R}^{\mathcal{T}}$
\State \textbf{Initialize}: Initialize reward matrix $O \in \mathbb{R}^{\mathcal{T} \times \mathcal{S}}$ with zero matrix. Initialize the prefix sum matrix $C \in \mathbb{R}^{\mathcal{T} \times \mathcal{S}}$ to the prefix sum of each row of $A$, that is, $C_{i,j} = \sum_{k=0}^{j}{[A]_{i,k}}$. Initialize all elements in the splitting boundary matrix $B_m \in \mathbb{R}^{\mathcal{T} \times \mathcal{S}} $ to zero.

\For{each $j\in [0,\mathcal{S})$}
    \State $[O]_{0,j} = [C]_{0,j}$
\EndFor

\For{each $i\in [1,\mathcal{T})$}
    \For{each $j\in [0,\mathcal{S})$}
        \For{each $k\in [0,\mathcal{S})$}
            \State $ O_{new} = [O]_{i-1,k} + [C]_{i, j}-[C]_{i, k}$
            \If{$ O_{new} > [O]_{i,j}$}
                \State $ [O]_{i,j} = O_{new} $
                \State $ [B_m]_{i,j} = k $
            \EndIf
        \EndFor
    \EndFor
\EndFor

\State $ P = \mathcal{S}-1 $

\For{each $i\in [\mathcal{T}-1, 0]$}
    \State $[D]_{i} = P - [B_m]_{i,P}$
    \State $P = [B_m]_{i,P}$
\EndFor

\State \textbf{return} $D$

\end{algorithmic}
\end{algorithm}

The duration extraction is shown as Algorithm~\ref{alg}. In Line 3, the element $[O]_{i,j}$ in reward matrix $O$ represents the maximum reward for the submatrix $[A]_{0...i,0...j}$, and the corresponding splitting point of the position $(i,j)$ is stored in $[B_m]_{i,j}$. From Line 4 to Line 17, we use DP to find a best spitting boundary and record the reward. From Line 18 to Line 22, we collect the best splitting boundary, starting from the position $(\mathcal{T}-1, \mathcal{S}-1)$ and tracing back, and in the meanwhile, calculate the duration of each phoneme.

\subsection{Details in Data Filtration}

We use the splitting reward introduced in Section~\ref{sec_alignment} to evaluate the alignment quality between singing and lyrics. We plot the alignments of some data and listen to these audios at the same time. We find that most alignments is accurate when the splitting reward $\mathcal{O}$ is larger than 0.6. Therefore, we set the threshold to 0.6 and filter singing data.

\subsection{Details in Singing Model}
\label{sec:append_singing}
\paragraph{Model Settings} We list the model settings of the singing model in DeepSinger in Table \ref{tab:hp_singing}.

\begin{table}[h]
\centering
\begin{tabular}{l|l}
\hline
\textbf{Model Setting}           &   \textbf{Value} \\ \hline
Phoneme Embedding Dimension       &     384   \\ \hline
Lyrics Encoder Layers             &     4     \\ \hline
Pitch Encoder Layers              &     2     \\ \hline
Reference Encoder Pre-net Layers  &     5     \\ \hline
Reference Encoder Pre-net Hidden  &     384   \\ \hline
Reference Encoder Layers          &     4     \\ \hline
Decoder Layers                    &     4     \\ \hline
Encoder/Decoder Hidden            &     384   \\ \hline
Encoder/Decoder Conv1D Kernel             &     3     \\ \hline
Encoder/Decoder Conv1D Filter Size        &     1536  \\ \hline
Encoder/Decoder Attention Heads           &     2     \\ \hline
Dropout                           &     0.1   \\ \hline \hline
Total Number of Parameters        &     34.22M  \\ \hline
\end{tabular}
\vspace{0.1cm}
\caption{Model settings of the singing model in DeepSinger. }
\label{tab:hp_singing}
\vspace{-0.8cm}
\end{table}

\paragraph{Exploration of Autoregressive Transformer-based Singing Model} 
We replace the decoder of our singing model with an autoregressive decoder which takes the last output of the decoder (linear-spectrogram frame) and the outputs of the encoder as inputs. In the inference process, we generate the linear-spectrograms in an autoregressive manner. The results show that the voice quality is not as good as our singing model with non-autoregressive generation. Since the error propagation problem in autoregressive generation under noisy data is much serious, non-autoregressive generation in our singing model shows advantages when giving accurate phoneme duration and pitch sequence.

\paragraph{Exploration of Mel-Spectrogram Auxiliary Loss} 
Inspired by Tacotron, which generates mel-spectrograms first and transforms the mel-spectrograms to linear-spectrograms, we also try to modify the singing model to make it generate mel-spectrograms first and then convert it to linear-spectrograms using a post-net. We minimize the MSE loss of mel-spectrograms and linear-spectrograms in the training stage and only use linear-spectrograms in the inference stage. The results show that the model trained with mel-spectrogram auxiliary loss does not gain any performance.

\subsection{Singing-Wild Dataset}
\label{sec:sw_detail}
\begin{figure}[!h]
    \centering
	\includegraphics[width=0.4\textwidth]{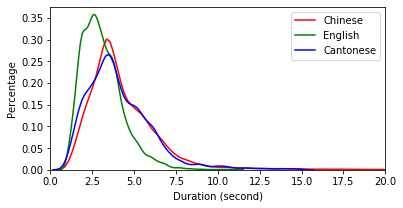}
    \vspace{-0.1cm}
	\caption{The distributions of sentence-level duration on three languages.}
	\label{fig_sent_duration}
    \vspace{-0.3cm}
\end{figure}

\begin{figure}[!h]
    \centering
	\includegraphics[width=0.4\textwidth]{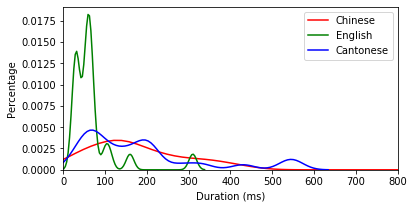}
	    \vspace{-0.1cm}
	\caption{The distributions of phoneme-level duration on three languages.}
    \label{fig_ph_duration}
    \vspace{-0.3cm}
\end{figure}

\begin{figure}[!h]
    \centering
	\includegraphics[width=0.4\textwidth]{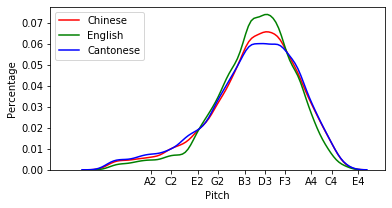}
    \vspace{-0.1cm}
    \caption{The pitch distributions on three languages.}
    \label{fig_pitch_dist}
    \vspace{-0.4cm}
\end{figure}

We plot the distributions of sentence-level duration, phoneme-level duration and pitch as analyzed in Section~\ref{sec_singing_wild} in Figure~\ref{fig_sent_duration}, \ref{fig_ph_duration} and \ref{fig_pitch_dist}.


\subsection{Evaluation Metric}
\label{sec:appendix_eval_metrics}

\paragraph{Mean Opinion Score (MOS)}
The scale of MOS is set to between 1 to 5. MOS from 1 to 5 denote bad, poor, fair, good and excellent respectively.

\paragraph{Preference Score}
We give each listener two voices and let the listener choose a better one or keep neutral based on the audio quality. 

\paragraph{Similarity Score} We use similarity score to evaluate how well the generated audio is similar to the singer in the reference audio. The similarity evaluation is performed by 20 people and they are told to focus on the similarity of the singer to one another rather than the content or audio quality of the audio. The scale of similarity score is set to between 1 to 5 which denote not at all similar, slightly similar, moderately similar, very similar and extremely similar respectively.

\subsection{Analysis on Cross-Lingual Synthesis}
\label{sec:appendix_xl}
We analyze whether DeepSinger can synthesize singing voices with the reference audio in one language but lyrics in another language, and evaluate the quality in terms of similarity score. For each language (English and Chinese), we conduct evaluation on the following settings: 1) the similarity between the ground truth audios in the same singer, which can be regarded as the upper bound (denoted as \textit{GT (Same)}); 2) the similarity between the ground truth audios from different singers, which can be regarded as the lower bound of similarity score (denoted as \textit{GT (Diff)}); 3) the similarity between the generated singing audios and reference audios in same language (denoted as \textit{Same Lan}); 4) the similarity between the generated singing audio and reference audio in different languages (denoted as \textit{Cross Lan}).

\begin{table}[!h]
	\centering
	\begin{tabular}{ l | c | c | c | c}
		\toprule
	Language  & \textit{GT (Same)} & \textit{GT (Diff)}  & \textit{Same Lan} & \textit{Cross Lan} \\
		\midrule
		 \textit{English}     & 4.53 $\pm$ 0.08  & 1.48 $\pm$ 0.13  & 3.39 $\pm$ 0.10 & 3.19 $\pm$ 0.14 \\
 		 \textit{Chinese}     & 4.65 $\pm$ 0.07  & 1.32 $\pm$ 0.10  & 3.68 $\pm$ 0.08 & 3.21 $\pm$ 0.17 \\
		\bottomrule
	\end{tabular}
	\vspace{0.1cm}
	\caption{The similarity scores of DeepSinger with 95\% confidence intervals.}
	\label{tab_sim}
    \vspace{-0.8cm}
\end{table}

We conduct analyses to synthesize voices on both Chinese and English to analyze the cross-lingual synthesis of DeepSinger.  As shown in Table~\ref{tab_sim}, DeepSinger achieves similarity scores above 3.0 (moderately similar) in ``Cross Lan'' setting, which demonstrates that the voice and timbre information captured by the reference encoder can be transferred across different languages.

\end{document}